\documentclass[fleqn,10pt]{wlscirep}
\usepackage[utf8]{inputenc}
\usepackage[T1]{fontenc}

\DeclareMathAlphabet{\mathcal}{OMS}{cmsy}{m}{n}
\DeclareSymbolFont{operators}   {OT1}{cmr} {m}{n}
\DeclareSymbolFont{letters}     {OML}{cmm} {m}{it}
\DeclareSymbolFont{symbols}     {OMS}{cmsy}{m}{n}
\DeclareSymbolFont{largesymbols}{OMX}{cmex}{m}{n}
\SetSymbolFont{operators}{bold}{OT1}{cmr} {bx}{n}
\SetSymbolFont{letters}  {bold}{OML}{cmm} {b}{it}
\SetSymbolFont{symbols}  {bold}{OMS}{cmsy}{b}{n}
\DeclareSymbolFontAlphabet{\mathrm}    {operators}
\DeclareSymbolFontAlphabet{\mathnormal}{letters}
\DeclareSymbolFontAlphabet{\mathcal}   {symbols}
\DeclareMathAlphabet      {\mathbf}{OT1}{cmr}{bx}{n}
\DeclareMathAlphabet      {\mathsf}{OT1}{cmss}{m}{n}
\DeclareMathAlphabet      {\mathit}{OT1}{cmr}{m}{it}
\DeclareMathAlphabet      {\mathtt}{OT1}{cmtt}{m}{n}

\usepackage{amssymb}
\usepackage{amsmath}
\usepackage{nameref}

\title{Spatial eco-evolutionary feedbacks mediate coexistence in prey-predator systems}

\author[1,*]{Eduardo H. Colombo}
\author[2]{Ricardo Mart\'inez-Garc\'{\i}a}
\author[1]{Crist\'obal L\'opez}
\author[1]{Emilio Hern\'andez-Garc\'{\i}a}
\affil[1]{IFISC (CSIC-UIB), Campus Universitat Illes Balears, 07122, Palma de Mallorca, Spain}
\affil[2]{Department of Ecology and Evolutionary Biology, Princeton University,  Princeton NJ 08544, USA}

\affil[*]{ecolombo@ifisc-uib-csic.es}

\begin{abstract}
Eco-evolutionary frameworks can explain certain features of communities in which ecological and evolutionary processes occur over comparable timescales. Here, we investigate whether an evolutionary dynamics may interact with the spatial structure of a prey-predator community in which both species show limited mobility and predator perceptual ranges are subject to natural selection. In these conditions, our results unveil an eco-evolutionary feedback between species spatial mixing and predators perceptual range: different levels of mixing select for different perceptual ranges, which in turn reshape the spatial distribution of prey and its interaction with predators. This emergent pattern of interspecific interactions feeds back to the efficiency of the various perceptual ranges, thus selecting for new ones. Finally, since prey-predator mixing is the key factor that regulates the intensity of predation, we explore the community-level implications of such feedback and show that it controls both coexistence times and species extinction probabilities.
\end{abstract}

\begin{document}

\flushbottom
\maketitle

\thispagestyle{empty}

% Use "Eq" instead of "Equation" for equation citations.
\section*{Introduction}

One of the major goals of ecology is to understand the mechanisms that sustain the coexistence of  antagonistic species, such as one prey and its predator, a host and its parasite, or multiple competitors for common resources. Under the traditional assumption that ecological and evolutionary changes occur on very different time scales, the connection between ecology and evolution is unidirectional, with the former driving the later. Therefore, the first attempts to explain species coexistence neglected the role of evolutionary processes and relied exclusively on ecological factors, such as species neutrality \cite{RevModPhys.88.035003}, frequency-dependent interactions \cite{ayala1971competition}, and environmental heterogeneity, either in space or in time~\cite{Chesson2000,Tarnita2015a,Martinez-Garcia2017,tilman1994competition,amarasekare2003competitive}.

More recently, however, evidences that ecological and evolutionary processes can occur at congruent time-scales have been found \cite{Siepielski2016,Kotil2018,Hiltunen2015}. This result suggests that both processes can affect each other and establish `eco-evolutionary feedbacks' (EEFs) that may alter the ecological dynamics and the stability of communities. Due to rapid evolution, the frequency of the genotypes and their associated phenotypes may change, within a population, as fast as ecological variables, such as population sizes or spatial distributions, and affect their dynamics. In turn, these new ecological configurations can redirect the evolutionary process \cite{naturalselection,ecoevo,Govaert2018,rapid,rapid2,Bonachela2017,mamunoz}.

The consequences of these EEFs at the community level have been studied mainly in single-species populations and simple two-species communities \cite{Govaert2018}. In prey-predator systems, empirical studies have shown that both prey and predator traits can evolve over ecological time scales, leading to EEFs that alter some features of the dynamics of both populations \cite{bohannan1999effect,Becks2010}. For instance, in a rotifer-algal system, rapid prey evolution induced by oscillatory predator abundance can drive antiphase in prey-predator cycles \cite{rapid}. Theoretical investigations have also suggested that prey-predator coevolution can induce a rich set of behaviors in population abundances, including reversion in the predator-prey cycles~\cite{Cortez2014}. Another family of studies has focused on the role of EEFs on the stability of the community, showing that different feedbacks influence the stability of prey-predator dynamics in different ways depending on the shape of the trade-offs between the evolving traits \cite{Abrams2001,Cortez2010,Govaert2018}.

However, despite these insightful studies, the interplay between eco-evolutionary feedbacks and spatial dynamics, this last being a crucial aspect that often controls species interactions, remains largely unexplored in ecological communities. EEFs in spatially structured populations have been studied mostly for single-species populations in which evolutionary dynamics affects the rate of dispersal, either across patches or during range expansion \cite{Govaert2018,Hanski2011,Fronhofer2017}. Here, we extend those scenarios and investigate how eco-evolutionary dynamics can modulate two-species interactions in a spatially-extended prey-predator community. To this aim, we use an individual-based model in which both species have limited mobility and only prey within a finite region around the predator are susceptible to predation. The radius of this region defines predators perceptual range, which in our model varies across the population and is subject to natural selection. Perceptual ranges, generally defined as the maximal distance at which individuals can identify elements of the landscape, vary tremendously within species and strongly determine the success of foraging and hunting strategies via several trade-offs~\cite{Zollner2000,Fagan2017}. For instance, large perceptual ranges increase the number of potentially detectable prey individuals, but may lead to a reduced attacking efficiency, as information is integrated over a large area~\cite{Martinez-Garcia2013,Martinez-Garcia2014,Fagan2017}. Moreover, the detection of many prey may also induce prey-crowding effects that reduce predation efficiency~\cite{lingle2001anti,vulinec1990collective}. These trade-offs bound the evolution of the perceptual range, setting a finite optimal value. Overall, due to its large intraspecific variability, important contribution to individual fitness, and sensitivity to species spatial distribution, the perceptual range arises as an important trait for studying the interplay between its evolutionary dynamics and spatial ecological processes within the community.

In fact, our results reveal that a feedback between the evolution of the predator perceptual range and species spatial distributions controls several community-level processes. We perform a systematic investigation of the prey-predator dynamics under different levels of mobility and mutation intensities and characterize the community long-time behavior using Shannon-entropy mixing measures, the distribution of predators' perceptual ranges, and species coexistence time and extinction probabilities. Depending on individual mobility (and the ecological interactions taking place), different levels of spatial mixing emerge, ranging from segregation to high mixing, and select for different perceptual ranges. Simultaneously, due to predation, perceptual ranges alter the spatial mixing of prey and predators, establishing an eco-evolutionary feedback. Importantly, since species mixing modulates the intensity of the prey-predator interaction,  the eco-evolutionary feedback strongly influences the stability of the community. A diagram summarizing the coupling between individual traits, species spatial distribution, and community-level processes is shown in Fig.~\ref{fig:ecoevo}.

Finally, although derived for the particular case of a prey-predator system, these results will more generally improve our understanding of how information gathering over different spatial scales may influence species interactions and how evolutionary processes may alter the ecological dynamics and stability of spatially-structured communities.

\section*{Results}

\begin{figure}[h]
\begin{center}
\vspace*{2mm}
\includegraphics[width=0.72\columnwidth]{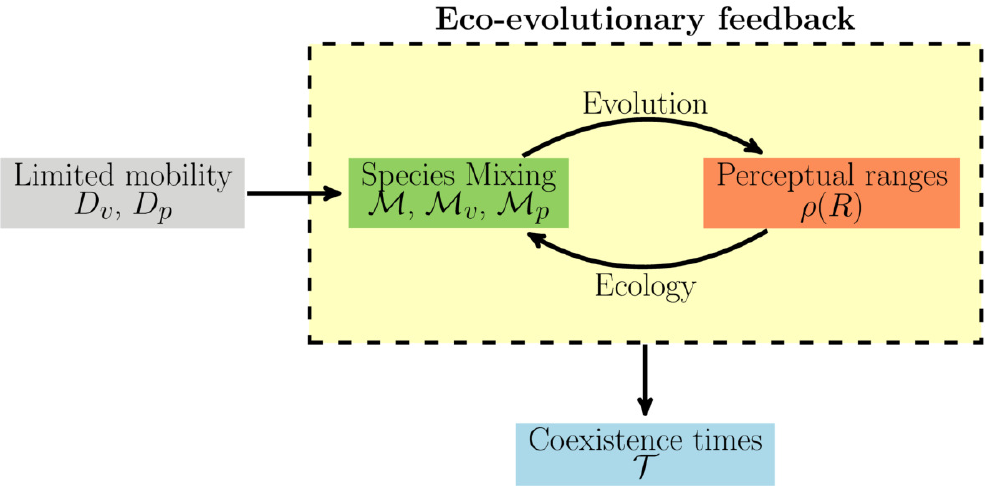}
\vspace*{4mm}
\caption{{\bf Schematic representation of the eco-evolutionary framework.} Diffusion coefficients, $D_v$ and $D_p$, control individual movement and act as control parameters of the emergent eco-evolutionary feedback (yellow box) between species mixing, measured using Shannon-entropy based metrics (green box), and predator perceptual ranges (orange box). Finally, the eco-evolutionary feedback determines prey-predator coexistence times $\mathcal{T}$. Arrows indicate the influence between the different elements of the framework.}
\label{fig:ecoevo}
\end{center}
\end{figure}

We aim to investigate the interplay between the spatial structure and the evolution of the spatial range of interspecific interactions in ecological communities (see the diagram in Fig.~\ref{fig:ecoevo}). To this end, we build an individual-based model for a simple prey-predator system (see Methods for full details) in
which individuals of both species move within a square environment of lateral length $L$ with periodic boundary
conditions. Movement is modeled using Brownian motions with diffusion coefficients $D_p$ and $D_v$  for predator and prey respectively ($v$ stands for victims). The intensity of the diffusion influences the spatial distribution of the populations. Strong diffusion leads to homogeneously distributed populations, whereas clusters form for weak diffusion due to the existence of reproductive pair correlations \cite{Young2001}.
% and to the interaction between the species \textbf{(I don't understand what does it mean ``interaction between species''. In Young 2001 there is not interaction between species and clusters still form for weak diffusion due to correlations in the parent-offspring positions)}.

The composition of the community is characterized by the number of prey individuals, $N_v=N_v(t)$, and predators, $N_p=N_p(t)$. These population sizes change in time driven by an asexual, stochastic population dynamics in which prey reproduction and predator death occur with constant rates $r$ and $d$ respectively. The predation or catching rate of each predator, $c$, however, is dictated by the availability of prey and the efficiency of the predator at attacking them. Mathematically, this can be written as $c(R)=E(R)M_v(R)$, where $M_v(R)$ accounts for the number of prey individuals within predator's perceptual range, $R$, and $E(R)$ is the attacking efficiency for a given perceptual range $R$. We have defined the perceptual range, different for each predator, as the maximum distance measured from the position of the predator at which a prey can be detected. Hence, the number of prey individuals detected by the predator increases as the perceptual range increases. We assume, however, that it does so at the cost of a reduced predation likelihood, leading to a trade-off between prey detection and  attacking efficiency. We implement this trade-off by assuming that the attacking efficiency $E(R)$ is a decreasing function of the perceptual range.  This trade-off between perception and attacking efficiency, as well as the choice of $E$ such that predation rate maximizes at intermediate scales of perception, is grounded on previous theoretical studies showing that foraging success decreases when individuals have to integrate information over very large spatial scales \cite{Martinez-Garcia2013,Martinez-Garcia2014,Martinez-Garcia2017b,Fagan2017,martinez2019range}.

\begin{figure}[h]
\begin{center}
\includegraphics[width=0.5\columnwidth]{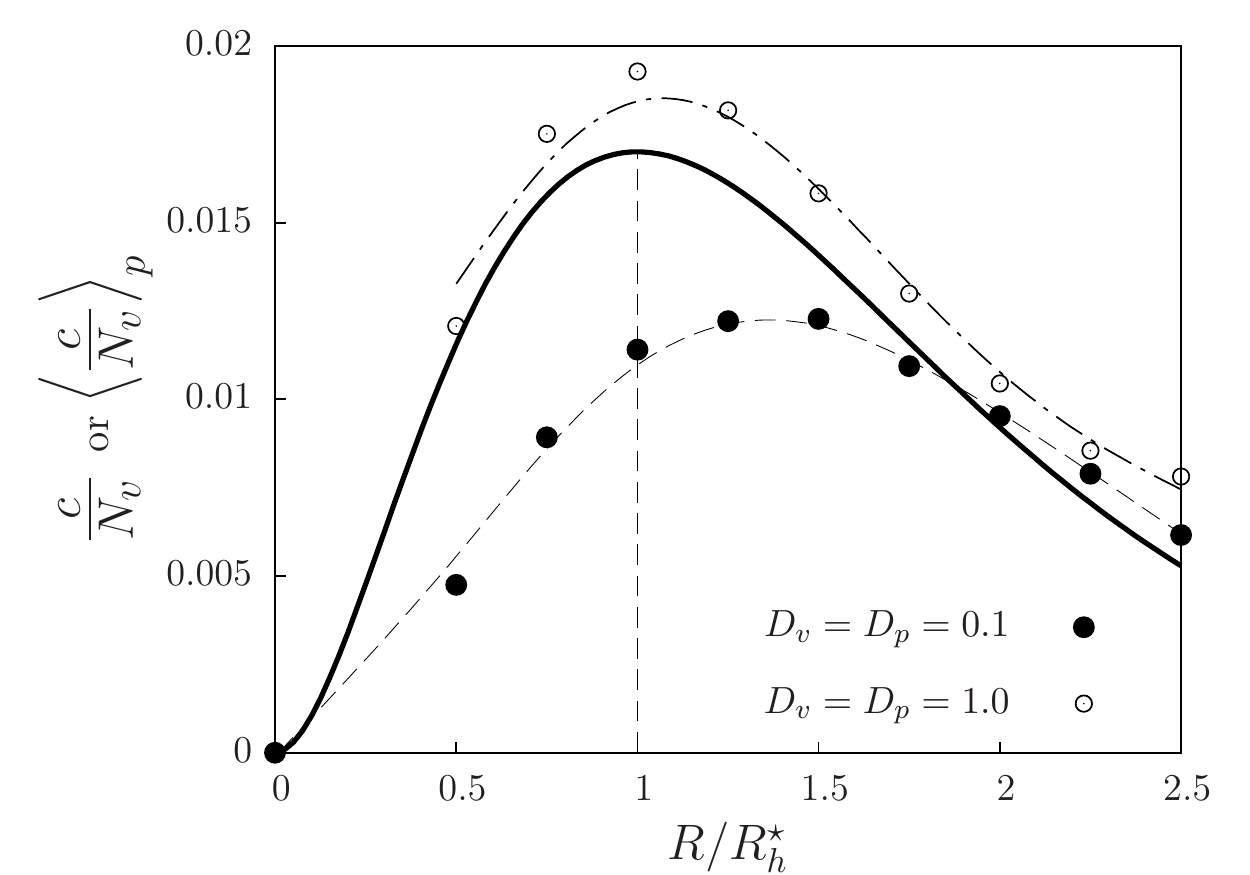}
\caption{{\bf Predation rate is maximum for intermediate perceptual ranges.} Solid line: predation rate per prey $c/N_v$ as a function of the predator's perceptual range, $R$, for the case in which prey individuals are homogeneously distributed in space (i.e. $c(R)/N_v = c_0 (\pi R^2/L^2)\exp(-R/R_c)$, see main text and Eq.~(\ref{catchingrate})). The trade-off between prey detection and decaying attack efficiency leads to a value $R=R^\star_h$ (indicated by a vertical dashed line) at which the predation rate per prey is maximum. Symbols (with dashed lines as guides to the eye) are, for each value of $R$, average predation rates per prey, $\left<c(R)/N_v\right>_p$ (average over all the predators in the system and over time and realizations), from simulations in which all predators have the same perceptual range $R$. There is no evolutionary process, and the mobilities are low ($D_v=D_p=0.1$, filled circles) or high ($D_v=D_p=1.0$, empty circles). Other parameters $r=d=c_0=b=R_c=1$, $L=10$.} \label{fig:catchingrate}
\end{center}
\end{figure}

The specific shape of $E(R)$ may depend on several factors related to prey behavior, predator behavior or environmental features. Here, we assume that it decays exponentially with the perceptual range as $E(R)=c_0\exp(-R/R_c)$, where $c_0$ is the maximum efficiency and $R_c$ fixes how quickly this efficiency decays as the perceptual range increases. We expect similar results for other functional forms of $E(R)$ as far as its decay with $R$ is faster than the growth of $M_v(R)$. Considering a homogeneous distribution of prey with density $v=N_v/L^2$, the number of prey detected by one predator is $M_v(R) = v \pi R^2$. Then, the predation rate for that predator is
\begin{equation}
c(R) = c_0 N_v \frac{\pi R^2}{L^2} \exp\left(-\frac{R}{R_c}\right),
\end{equation}
which is maximal for $R^\star_h \equiv 2 R_c$ (see the solid line in Fig.~\ref{fig:catchingrate} for a plot of $c(R)/N_v$, normalized to make it independent of the prey density). 
In the next sections,  we will use this value $R^\star_h$ as a reference to measure the effect of non-homogeneous distributions of individuals on the optimal perceptual range. %Note that we do not model the attack process, that we consider to be instantaneous. Thus, the predator mobility described by the diffusion coefficient $D_p$ refers to the predator motion while searching.
When the spatial distributions of the populations are heterogeneous, the predation rate per prey varies across the population because the number of prey perceived by each predator does so. In these cases, we measure the mean predation rate per prey, $\left<c(R)/N_v\right>_p$, from numerical simulations of the non-evolutionary dynamics in which all predators have the same perceptual range $R$ and there are no mutations. 
The notation $\left<...\right>_p$ indicates average over all predators in the system and over time and realizations. 
If individual diffusion is low, the spatial distribution of individuals deviates strongly from homogeneity, thereby leading to optimal perceptual ranges,  $R^*$, that are larger than in the well-mixed limit (see filled symbols in Fig.~\ref{fig:catchingrate}). Conversely, as diffusion increases, $M_v(R) \to N_v \pi R^2/L^2$ and the system approaches the well-mixed limit (see empty symbols in Fig.~\ref{fig:catchingrate}). In this high-diffusion limit, however, small differences with the analytical result remain because our simulations consider finite populations that always show some spatial heterogeneity.% We have checked that the small offset between solid line and empty circles disappears in situations with larger number of individuals in the system, indicating that it is a consequence of the corrections to homogeneity intrinsic to a finite number of discrete individuals, even if they are very well mixed. 
%In fact, the number ofbpreys within predators reach can be written as $M_v/v = 2 \pib\int_0^R g(r) r dr$, where $g$ is the pair correlation functionvbetween predators and preys, which becomes $g(r)=1$ for thevwell-mixed (uncorrelated) case.
%Latter, we will use Shannon-entropy mixing measures to quantify deviations from spatially homogeneous distributions of individuals, and to relate them to community-level characteristics of our system.

Finally, once a prey is consumed by a predator with the individual-specific predation rates defined above, there is a probability $b$ for the predator to reproduce. Hence, predator's reproduction rates are determined by the interplay between their perceptual range and the spatial distribution of the prey population. Neglecting the phenotype-genotype distinction and the role of the environment in trait inheritance~\cite{Bird2007}, we assume that each newborn predator receives the perceptual range from its parent (i.e. clonal reproduction) plus some mutation that adds to $R$ a random perturbation sampled from a Gaussian of zero mean and variance $\sigma^2_\mu$. $R$ remains unchanged during predators' lifetime. The intensity of mutations, $\sigma_\mu$, determines both the speed of the evolutionary process and the level of variability in the evolving trait, $R$. The mathematical details of the model and its implementation are provided in the Methods section.

In the following sections, we investigate how the coupling between limited mobility and evolution in the perceptual ranges influences the stability of the community. Hence, we keep constant all the model parameters (see Methods section) except the intensity of the mutations in $R$, $\sigma_\mu$, that drives the evolutionary processes, and the diffusion coefficients, $D_v$ and $D_p$, which determine the degree of mixing in the population and thus the intensity of the prey-predator interaction.
%(for computational convenience, different values of $L$ will also be used). 
Therefore, for a given pair of diffusion coefficients and a mutation intensity, three linked population and community-level features emerge: the spatial distribution of both species, the distribution of predator perceptual ranges (i.e., the outcome of the evolutionary dynamics) and the coexistence time of the populations. In the following sections, we study each of them separately.

%Note that if we switch-on now the evolutionary dynamics (by introducing mutations with $\sigma_\mu\neq 0$ or by initializing the predators with different values of $R$, so that selection can take place) the perceptual-range distribution $\rho(R)$ is not a free parameter in our model. As we will see, the most frequent perceptual range in the predator population $R^\star$ (the mode of $\rho$) will be a function of the species diffusion coefficients. Along the next sections, we investigate and address how the system behaves as a function of individual mobility, highlighting the role of evolutionary dynamics in the community outcomes.

\begin{figure}[h]
\begin{center}
\includegraphics{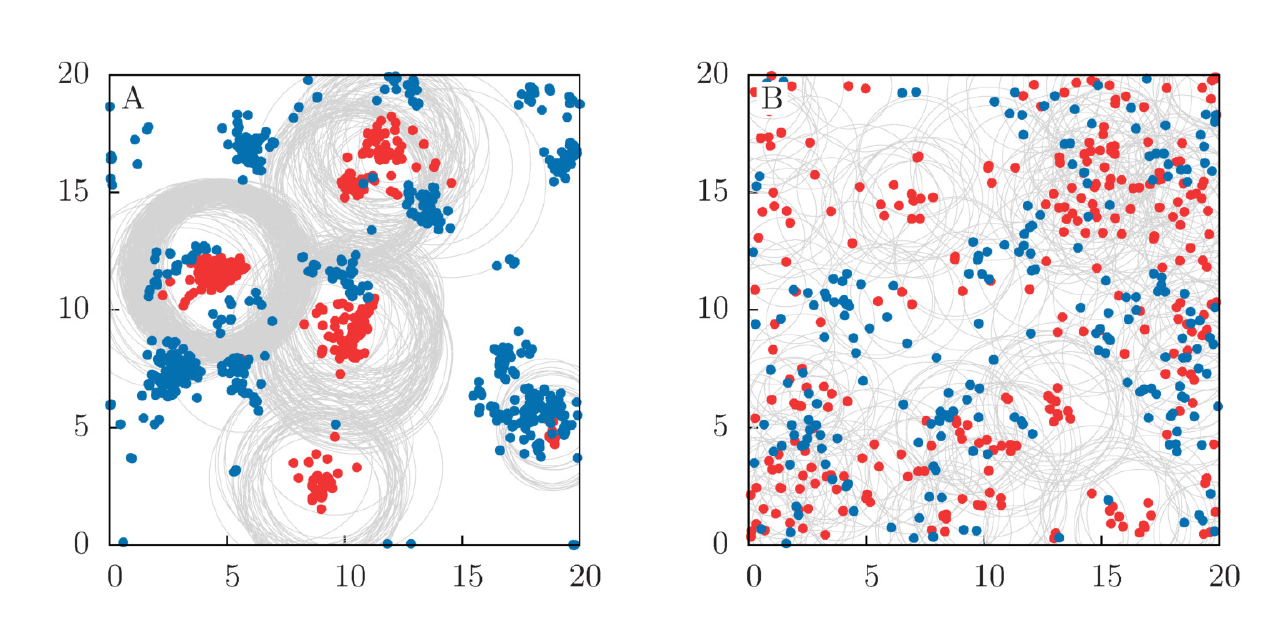}
\caption{{\bf Species spatial distribution.} Spatial distribution of predators (red) and prey (blue) in the
long-time regime for (A)
low and (B) high mobility, with $D_p=D_v=0.1$ and $D_p=D_v=1$, respectively. Gray circles
delimits the area of perception of the predators, which is subject to  evolutionary
dynamics ($\sigma_\mu=0.1$, see Methods section for details).
The habitat is a square domain with size $L=20$ and periodic
boundary conditions (i.e. our habitat is actually the surface of a torus).
See movies in the Supplementary Material to visualize the model dynamics.}
\label{fig:distribution}
\end{center}
\end{figure}

\subsection*{Species spatial distributions}

If prey birth and predator death rates are kept constant, the spatial distributions of prey and predators is determined by three characteristic spatial scales, defined by $D_v$, $D_p$ and $R$. Figure \ref{fig:distribution} shows typical spatial configurations obtained for low (panel A) and high (panel B) diffusion (see also movies in the Supplementary Material). Clustering of individuals and species segregation occur at low diffusion. To quantify population clustering within each species, as well as interspecies mixing, we introduce the metrics $\mathcal{M}_v$ and $\mathcal{M}_p$ for the former and $\mathcal{M}$  for the latter. The three of them are defined in terms of the Shannon index or entropy~\cite{diversityEntropy,entropyindex,cristobalentropy}, conveniently modified to correct for the effect of fluctuations in population sizes (see Methods for the mathematical definitions and further details). The interspecies mixing, $\mathcal{M}$, takes values between $0$ and $1$, with $0$ indicating strong species segregation and $1$ the well-mixed limit. $\mathcal{M}_p$ and $\mathcal{M}_v$ also take values within the same range, but since they are applied to one single species, $\mathcal{M}_\alpha=0$ indicates a high level of clumping of species $\alpha$ ($=v$ or $p$) and $\mathcal{M}_\alpha=1$ a uniform distribution of the corresponding species.

\begin{figure}[h]
\begin{center}
\includegraphics{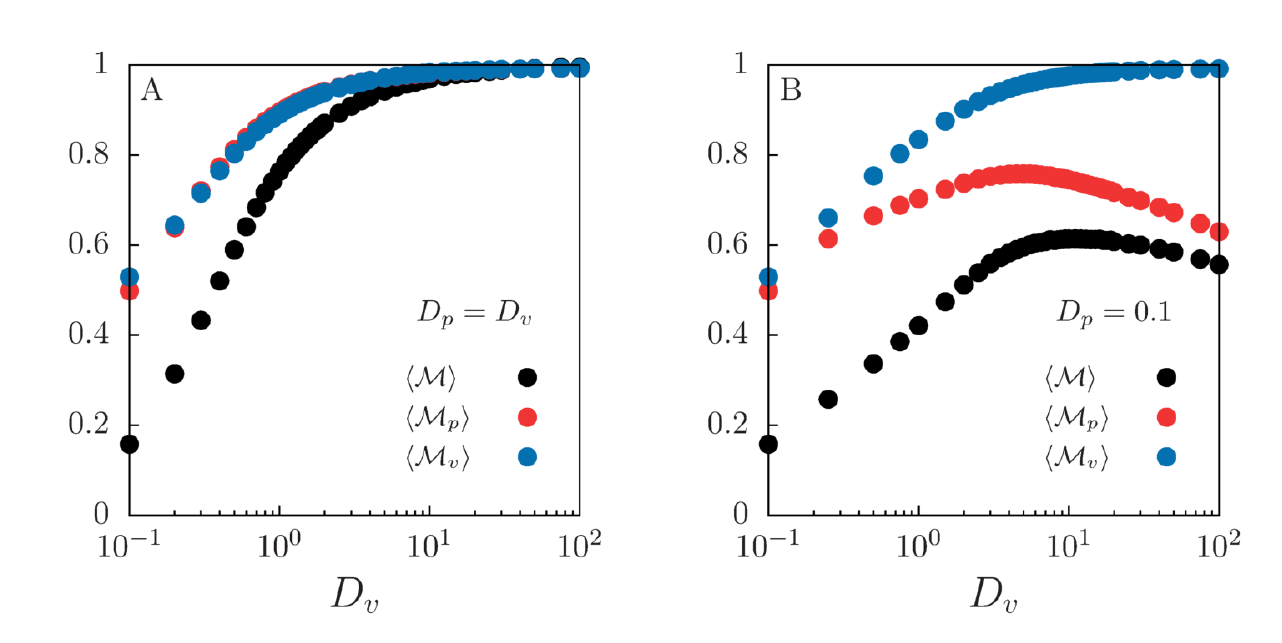}
\caption{{\bf Characterization of the spatial distribution through Shannon-entropy mixing measures.} Average prey-predator mixing $\langle \mathcal{M} \rangle$ and prey and predator mixing, $\langle \mathcal{M}_v\rangle$ and $\langle \mathcal{M}_p \rangle$ respectively, for different individual's mobility with (A) $D_p=D_v$ and (B) $D_p=0.1$. Mutation intensity is $\sigma_\mu=0.1$ and habitat size $L=10$. Averages are performed over time and $10^4$ independent realizations in the long-time regime.}
\label{fig:mixing}
\end{center}
\end{figure}

We analyze the spatial distribution of species in the long-time regime, that is, once the distribution of perceptual ranges in the predator population has reached a stationary shape. In Fig.~\ref{fig:mixing}, we show the mean value (the notation $\left<...\right>$ indicates average over time and realizations) of these mixing measures as a function of individual mobility for constant mutation intensity $\sigma_\mu=0.1$. Our results reveal a complex interaction between mobility and species mixing. When both prey and predators have the same diffusion coefficient, $D_v=D_p$, all the mixing indexes increase with species diffusion (Fig.~\ref{fig:mixing}A). However, when prey and predator have different diffusion coefficients, $D_v\neq D_p$, some of the mixing indexes may become non-monotonic functions of one of the diffusion coefficients. For instance, in the particular case shown in Fig.~\ref{fig:mixing}B, prey mixing still increases monotonically with $D_v$, but both interspecies and predator mixing show a maximum at intermediate $D_v$. The prey population can be seen as a dynamical resource landscape that drives the spatial distribution of predators. Increasing $D_v$ always leads to a more uniform distribution of prey. However, the extent to which this also leads to a more uniform distribution of predators is limited by $D_p$ (which in Fig.~\ref{fig:mixing}B is kept constant and at a low value $D_p= 0.1$). For instance, when $D_v \gg D_p$ predators cannot follow the dynamics of the prey and both $\mathcal{M}$ and $\mathcal{M}_p$ decrease with increasing $D_v$.

\begin{figure}[htb]
\begin{center}
\includegraphics[scale=1]{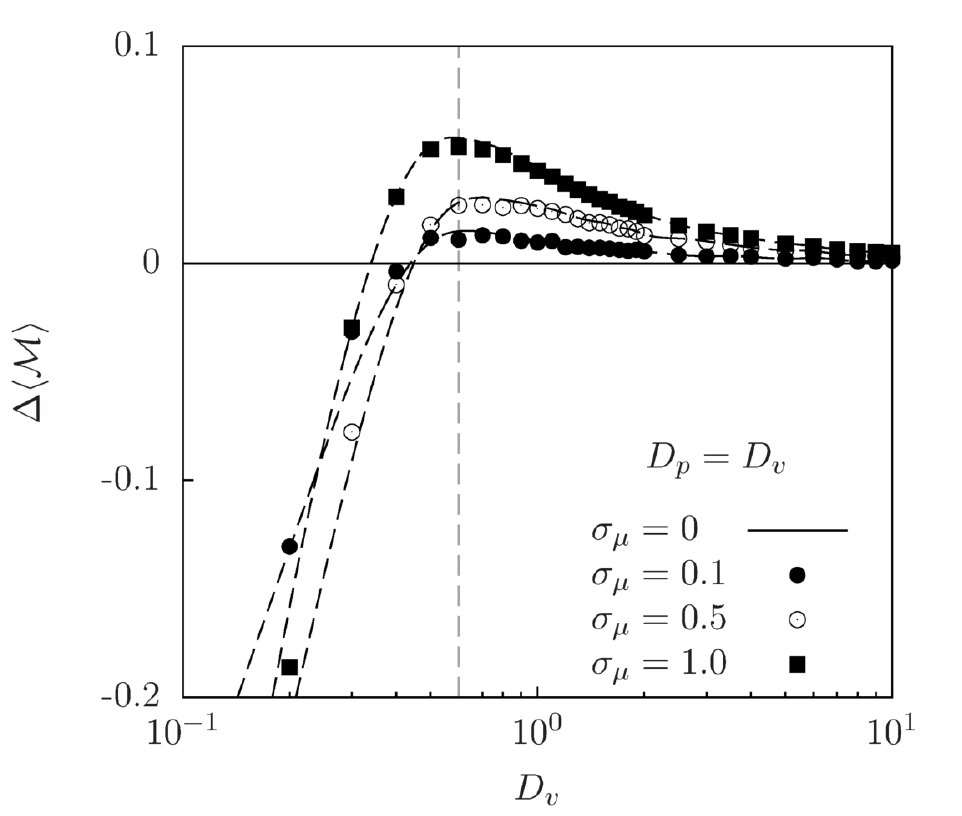}
\caption{{\bf Effect of mutation intensity and diffusion on population mixing.} Change in the average prey-predator mixing relative to the no-mutation case, $\Delta \langle \mathcal{M} \rangle$ (Eq.~(\ref{deltamix})), as a function of the diffusion coefficients $D_v=D_p$. Different symbols represent different levels of mutation intensity. Habitat size $L=10$. Symbols are the results from numerical simulations and dashed lines are smooth fits to simulation data for different mutation intensities. The horizontal continuous line $\Delta \langle \mathcal{M} \rangle=0$ is the no-mutation case ($\sigma_\mu=0$) and the vertical line indicates the maximum mixing for $\sigma_\mu=1.0$. }
\label{fig:deltamix}
\end{center}
\end{figure}

Next, we explore the effect of the mutation intensity on the average interspecies mixing, $\langle \mathcal{M}\rangle$. In Fig.~\ref{fig:deltamix}, keeping $D_v=D_p$ constant, we show how $\sigma_\mu$ changes the prey-predator mixing curve shown in Fig.~\ref{fig:mixing}A. To quantify such change, we define the relative change in $ \langle \mathcal{M} \rangle $ with respect to a no-mutation case ($\sigma_\mu\to 0$) in which all predators have the optimal perceptual range $R^\star$,
\begin{equation}
\label{deltamix}
\Delta \langle \mathcal{M} \rangle (D_v,D_p| \sigma_\mu) \equiv  \frac{\langle\mathcal{M}\rangle(D_v,D_p | \sigma_\mu) - \langle\mathcal{M}\rangle(D_v,D_p | \sigma_\mu = 0) }{\langle\mathcal{M}\rangle(D_v,D_p | \sigma_\mu=0)},
\end{equation}
where brackets indicate an average performed over time and realizations in the long-time regime. We find that, at low diffusion, interspecies mixing decreases as mutation intensity $\sigma_\mu$ increases. This leads to more segregated prey-predator distributions for larger mutation intensities. At high diffusion, however, the trend is reversed and interspecies mixing increases with increasing mutation intensity. Finally, for a range of intermediate diffusion coefficients, the increase in interspecies mixing due to mutation is maximum. These results arise mainly from the mutation-induced variability in the values of $R$ within the predator population, which will be discussed in the next section.

\subsection*{Distribution of predator perceptual ranges}
\label{sec:evolution}

In our model, we assume that predator perceptual ranges (and thus predation rates and predator reproduction rates) are subject to natural selection. We assume that the value of the trait $R$ of a predator is passed to its offspring, with some variation due to mutation. Then, the perceptual range remains unchanged during the individual lifetime. Natural selection is at work since, depending on the spatial distribution of prey, some perceptual ranges are favored against the others and hence tend to be over-represented within the population.

\begin{figure}[h]
\begin{center}
\includegraphics{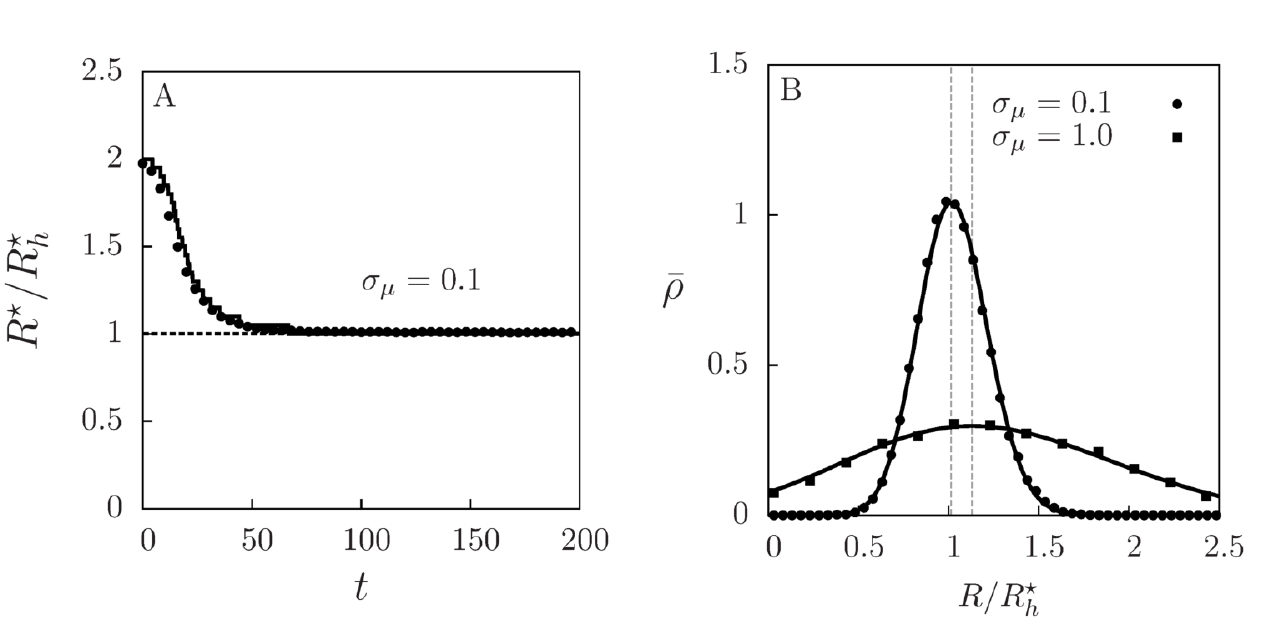}
\caption{{\bf Evolutionary dynamics in the homogeneous limit.} (A) Temporal evolution of the dominant perceptual range $R^*$ (the mode, i.e. the maximum of $\rho(R)$), relative to the one giving the maximum predator growth in the homogeneous case, $R^*_h$, for $\sigma_\mu=0.1$ and a system size $L=40$. Solid line is obtained from the numerical solution of Eqs.~(S2)-(S3) derived in the Supplementary Material. Dots correspond to numerical simulations of the individual-based model with $D_v,D_p\rightarrow\infty$, obtained by averaging the distribution of perceptual ranges over $100$ independent runs and then extracting its maximum $R^\star$ at every time. In all cases, the initial distribution of perceptual ranges is sharply peaked at $R=2R_h^\star$. (B) Probability density for finding a perceptual range value $R$ in the population of predators, $\bar \rho(R)=\rho(R)/N_p$, in the long-time regime, for low and high mutation intensity. Dots correspond to simulations of the individual-based model (with $D_v,D_p\rightarrow\infty$, average over 100 runs) and solid lines to the theoretical prediction (see Supplementary Material for details). Dashed vertical lines show the position of the mode of each distribution.}
\label{fig:HOMcase}
\end{center}
\end{figure}

\textbf{Spatially homogeneous limit.} In the $D_v$, $D_p\rightarrow\infty$ limit (which leads to $\mathcal{M},\mathcal{M}_v, \mathcal{M}_p\to 1$), the populations of prey and predators are randomly distributed in space and well-mixed with each other. In this homogeneous, mean-field limit, it is possible to derive an equation for the dynamics of the distribution of perceptual ranges in the population, $\rho(R)$ (see Supplementary Material for detailed calculations).  In this limit, we can approximate the expected number of prey individuals within a radius $R$ by $M_v(R) \approx \pi R^2 v$, where $v$ is the (uniform) prey density. Thus, the predation rate is $c(R) = E(R) M_v(R) \approx c_0 \pi R^2 v e^{-R/R_c}$. As long as the number of individuals used in the simulations is large, this theoretical prediction (see Eqs.~(S2) and (S3) in the Supplementary Material) agrees with direct simulations of the individual-based dynamics (see Fig.~\ref{fig:HOMcase}). The infinite-diffusion, well-mixed limit is implemented in the simulations by randomly redistributing prey and predators in space at each time step. Starting from different initial distributions for $R$, the maximum of the time-dependent distribution $\rho(R)$, which defines the dominant perceptual range $R^\star$, is driven close $R_h^\star$, that is, the perception range that gives the maximum predation rate in the homogeneous regime (Fig.~\ref{fig:HOMcase}A). This long-time dominant value corresponds exactly to the optimal one, $R_h^\star$, when mutations are negligible. However, as the intensity of mutations, $\sigma_\mu$, increases, the long-time distribution $\rho(R)$ becomes wider and its mode shifts towards larger values of $R$ (Fig.~\ref{fig:HOMcase}B). This effect is due to the asymmetric form of the predation rate $c(R)$ (Fig.~\ref{fig:catchingrate}). Thus, the dominant $R$ in the homogeneous regime has a main component set by the optimal value and a small positive shift due to the effect of mutations. Lastly, note that because the diffusion coefficients are very high, any perturbation of the homogeneous spatial distribution is rapidly smoothed out. Thus, selection for a specific $\rho(R)$ driven by the evolutionary dynamics has no effect on the spatial distribution of species. This is in contrast with the result for low mobility, which we discuss more in depth next.

\begin{figure}[h]
\begin{center}
\includegraphics[width=\columnwidth]{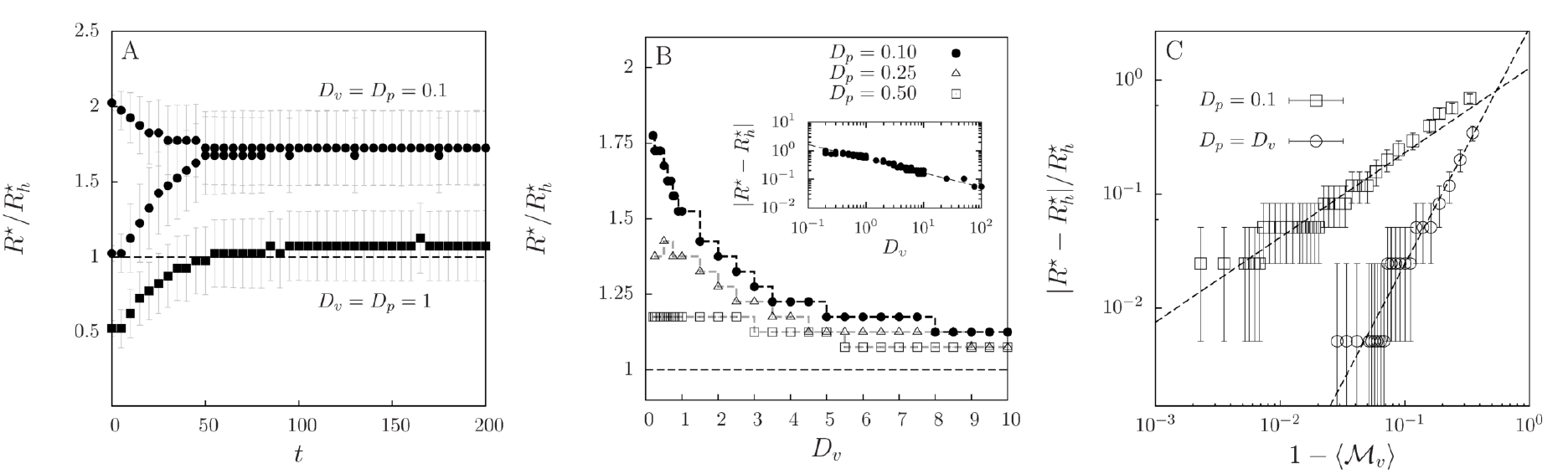}
\caption{{\bf Dominant perceptual range: from the segregated to the well-mixed scenario.} (A) Temporal evolution of the location $R^\star$ of the maximum in the average
perceptual-range distribution $\rho(R)$ (average over $10^4$ runs), relative to the optimal
perceptual range for homogeneous populations, $R^\star_h$, for high ($D_v=D_p=1.0$) and
low ($D_v=D_p=0.1$) diffusion coefficients. Two different sharply-localized initial
population distributions $\rho(R)$ are used in each case. Bars indicate the standard deviation of $\rho(R)$ around $R^\star$.
(B) Dominant perceptual range relative to the optimal perceptual range in the well-mixed limit,
$R^\star/R^\star_h$, as a function of prey and predator diffusion rates.
Dashed lines are guides to the eye and display the discretization $\Delta R= 0.1$ used for
the numerically obtained $\rho$. Inset shows the asymptotic approach of $R^\star$ to $R^*_h$ as $D_v$ increases with $D_p=0.1$.
(C) The relative difference between the dominant perceptual range in the simulations
and the optimal one in the homogeneous case,
$|R^\star - R^\star_h|/R^\star_h$ versus $1-\langle{\cal M}_v\rangle$, which measures
prey clumping (averaged over time and realizations in the long-time regime). Each symbol correspond a prey diffusion $D_v$ (ranging from $10^{-1}$ to $10^2$), while keeping $D_p=D_v$ (circles) or
$D_p=0.1$ (squares). Bars indicate
bin size of the computationally obtained $\rho(R)$. Dashed lines represent
the power-law expressions set in Eq.~(\ref{powerlaw}), with $\gamma\simeq 1.5$ for $D_p=D_v$ and
$\gamma\simeq 0.5$ for $D_p=0.1$. Habitat lateral length $L=10$ and mutation intensity $\sigma_\mu=0.1$ in all the panels.}
\label{fig:evo-comp}
\end{center}
\end{figure}

\textbf{Finite-mixing case.} For the general case of limited dispersal, far from the well-mixed scenario,  some of the features shown in Fig.~\ref{fig:HOMcase} still persist, but modified due to the underlying spatial distribution of prey and predators. Since the analytical approximations derived for the infinite diffusion limit are not valid, we study this scenario via numerical simulations of the individual-based model. Starting from different initial distributions of $R$, the most frequent (probable) value of $R$, $R^\star$, evolves in time towards a value that depends on the mobility of both species (Fig.~\ref{fig:evo-comp}A), with lower mobility favoring larger ($R^\star > R_h^\star$) perceptual ranges (Fig.~\ref{fig:evo-comp}B), while, in the well-mixed limit, it approaches the $R_h^\star$ as a power-law (see inset of Fig.~\ref{fig:evo-comp}B). The change in the long-time $\rho(R)$, both with time and with diffusion coefficients, is shown in the supplementary figure Fig. S1. 
We observe that the change in the dominant perceptual range due to species diffusion is well captured by the prey mixing parameter $\mathcal{M}_v$: Fig.~\ref{fig:evo-comp}C shows the dominant $R$ in the long-time regime as a function of prey clumping generated varying $D_v$. We extract that
\begin{equation}
\label{powerlaw}
\frac{|R^\star -R^\star_h|}{R_h^\star} = (1-\langle \mathcal{M}_v \rangle)^{\gamma}\, ,
\end{equation}
with $\gamma\simeq 1.5$ ($\gamma\simeq 0.5$) when fixing $D_p=D_v$ ($D_p=0.1$) and mutation intensity $\sigma_\mu=0.1$. This relation is valid for low mutation intensity, such that in the well-mixed scenario, $\mathcal{M}_v\to 1$ (achieved for large diffusivities), we have $R^\star \simeq R^\star_h$. Prey mixing is thus the main quantity that determines the dominant perceptual range, because it captures the major contribution of the spatial structure of the landscape of resources experienced by predators. As prey form clusters, $\mathcal{M}_v < 1$, predators typically find prey at distances that are larger than in the homogeneous case (see Fig.~\ref{fig:distribution}), which drives the evolution of $R^\star$ to larger values.

% I COMMENT ALL THIS PARAGRAPH SINCE I DO NOT UNDERSTAND VERY WELL NOR SEE ITS RELEVANCE
%We also note that the stationary dominant trait $R^\star$ is sensitive
%to the mutation noise intensity. For, all the above results we used that $\sigma_\mu=0.1$,
%in which this change is negligible, however for larger mutation noise intensities this change is notable,
%feature captured by Eq.~(\ref{macroevo}) as shown in Fig.~\ref{fig:HOMcase}.
%This introduces a small positive shift in the curves shown Fig.~\ref{fig:evo-comp}, keeping the overall picture unchanged.
% The shift observed in the homogeneous case is the same to the ones observed for the general cases investigated
% in Fig.~\ref{fig:optimalDiff}, then it do not affect the law found in Eq.~(\ref{fig:optimalMix}).

\subsection*{Coexistence times and extinction probabilities}
\label{sec:coexistence}

We have investigated in the previous sections the mutual influence between species mixing and the evolution of the perceptual-ranges. Thus, since mixing controls the frequency of prey-predator interactions, we expect this interplay between ecological and evolutionary processes to mediate the stability of the community. To quantify this, we measure the mean coexistence time between prey and predators, $\mathcal{T}$, and the probability that prey get extinct before predators, $\beta$, as a function of prey diffusion, predator diffusion, and the intensity of the mutations. The mean coexistence time, $\mathcal{T}$, is defined as the time until either prey or predators get extinct, averaged over independent model realizations. Prey extinction probability, $\beta$, is obtained as the fraction of realizations in which predators persist longer than prey. Since prey are the only resource for predators, predators will shortly get extinct following prey extinctions. On the contrary, when predator extinctions occur first, prey will grow unboundedly because we do not account for interspecific competition. For each realization, we use initial conditions that lead to a very short transient after which the spatial structure of the populations and the perceptual-range distribution achieve their long-time shape. In most cases, an initial condition consisting of spatially well-mixed populations and a uniform distribution of perceptual ranges in $[0,L/2]$ allows this to happen. Nevertheless, for small mutation rates ($\sigma_\mu<0.1$), the evolutionary time scales become comparable to the coexistence times and we need to speed up the evolutionary component of the transient dynamics by setting an initial condition for the perceptual-range distribution close to the one expected at long times.

\begin{figure}[htb]
\begin{center}
\hspace{-1cm}
\includegraphics{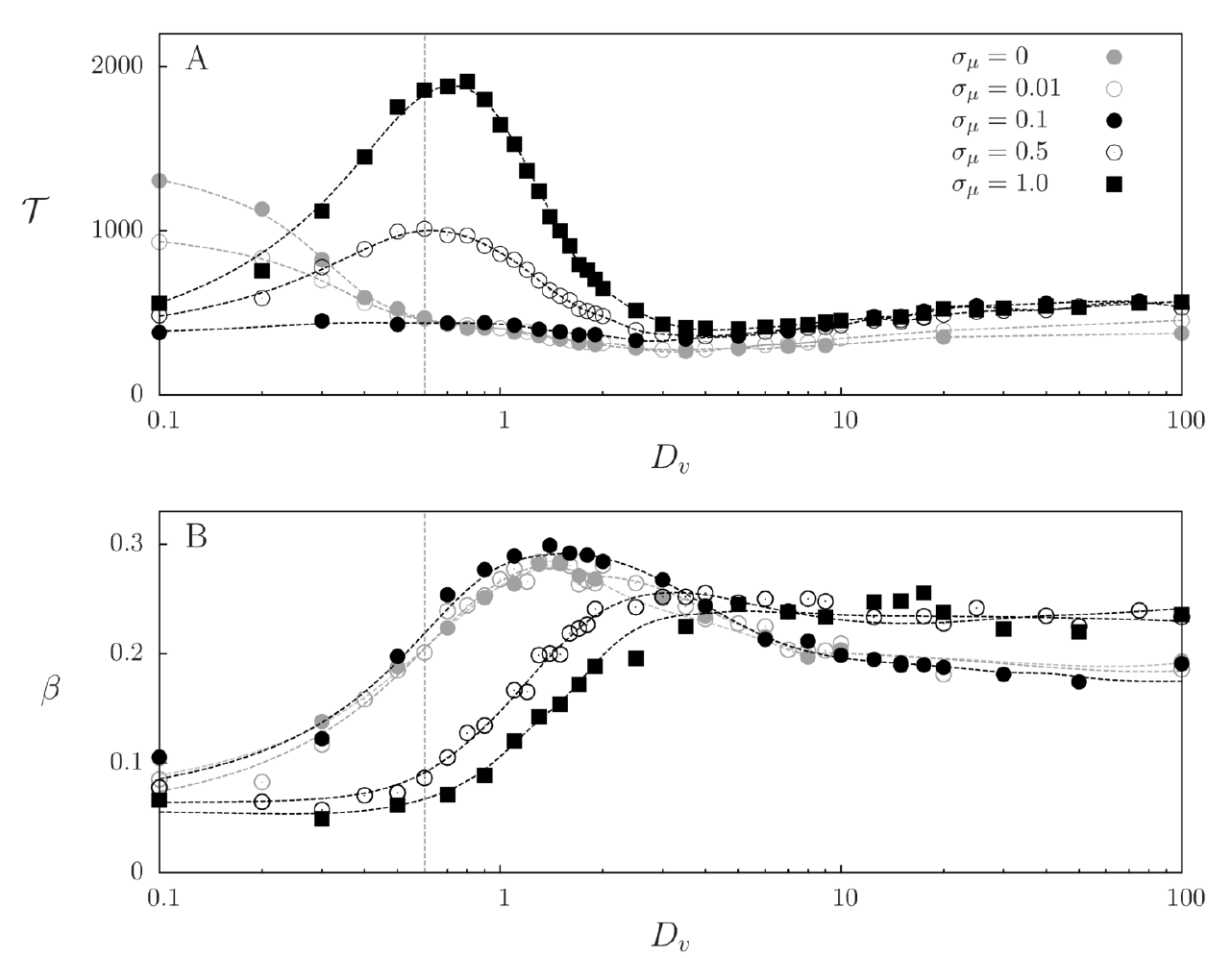}
\vspace{0.5cm}
\caption{{\bf Community coexistence times and prey extinction probability.} (A) Mean coexistence time $\mathcal{T}$ and (B) probability of prey extinction before predators, $\beta$, as function of the diffusion coefficients $D_p=D_v$ for different levels of mutation noise intensities with system size $L=10$. Initial conditions are prey and predators uniformly distributed in space with perceptual range $R$ uniformly distributed in $[0,L/2]$, and results were extracted from $5\, \times \,10^3$ realizations. Dashed lines are smooth fits to guide the eye. Vertical dashed lines indicate the diffusivity value at which the increase of mixing with respect the no-mutation case is maximum (from Fig.~\ref{fig:deltamix}, $\sigma_\mu=1.0$)} \label{fig:coexistence}
\end{center}
\end{figure}

In Fig.~\ref{fig:coexistence}A we show the mean coexistence time $\mathcal{T}$ as a function of prey and predator diffusion coefficients, assuming $D_v=D_p$, for different mutation intensities. This curve depends, in a complex manner, on the values of the dominant perceptual range,  the associated predation rate, and the degree of mixing that arises from limited dispersal. Long coexistence occurs when there is a balanced mixing between prey and predators, which results in intermediate levels of predation that preserve prey longer. For weak mutation, the coexistence time, which is maximum at low diffusion, decreases as the diffusion coefficients increase until reaching a minimum at intermediate mobility. Then, $\mathcal{T}$ increases slowly, approaching asymptotically the well-mixed case. As mutation increases, there is a clear change in the dependence of $\mathcal{T}$ with the diffusion coefficients. Beyond a value of $\sigma_\mu$, the maximum of $\mathcal{T}$ shifts to intermediate values of the diffusion coefficients. This is one of the central results of this paper, coming from the effect that mutation intensity (when non-negligible) has in prey-predator mixing, as shown in Fig.~\ref{fig:deltamix}: mixing decreases for low diffusion and increases until intermediate values of diffusion coefficients. Since mixing controls interspecies interaction, a key ingredient for coexistence, this is translated to the behavior of $\mathcal{T}$. As seen in Fig.~\ref{fig:coexistence}A, the level of mobility at which the increase in mixing is maximum (vertical dashed line, from Fig.~\ref{fig:deltamix}) roughly matches the location of the maximum $\mathcal{T}$.

Finally, we calculate the probability that prey become extinct before predators, $\beta$, as a function of $D_v$ (which is taken to be equal to $D_p$) for different values of $\sigma_\mu$ (Fig.~\ref{fig:coexistence}B). Even though the most likely event is that predators disappear before prey ($\beta<0.5$), as the diffusion coefficients increase from very small values, we observe an increase on $\beta$ passing through a maximum at intermediate diffusion. Despite the nonlinear effects between predation rate and species spatial distributions, spatial mixing enhances predation and therefore $\beta$ generally becomes larger as diffusion increases (see Fig.~\ref{fig:coexistence}B). Note that, comparing Figs.~\ref{fig:coexistence}A and~\ref{fig:coexistence}B, the maximum $\beta$ (high predation) is not related to longest coexistence, which indicates that species coexist longer when there is a balance between predation and prey reproduction. The influence of the intensity of mutations in the profiles shown in Fig.~\ref{fig:coexistence}B, again, is due to the feedback in the interspecies mixing shown in Fig.~\ref{fig:deltamix}, which regulates the level of predation. Hence, prey extinction is reduced at low mobility but increased at high mobility, shifting the profile.

\subsection*{Relationship with the non-evolutionary case}
\label{sec:nonevo}

In order to better understand the origin of the spatial eco-evolutionary feedback and its importance to our results, we have thoroughly explored also a non-evolutionary simulation scenario in which all predators have the same, fixed value of $R$. We have scanned the $R-D_v$ parameter space (with $D_v=D_p$) and measured the behavior of the different population and community-level properties studied in previous sections (spatial mixing, average predation rate, mean coexistence times, and prey extinction probability). The results, shown in Fig.~S2, summarize the consequences of the eco-evolutionary process, specially at low mutation intensity.  In this mutation-intensity regime, the evolutionary dynamics generates, when all the rest of system parameters remain constant, a narrow distribution of perceptual ranges that singles out a preferred $R^\star$. Thus, as the individual mobility is varied, the optimal perceptual range (driven by evolution) follows a specific curve $R^\star(D_v)$ in the $R-D_v$ plane (the black solid line in Fig.~S2) and the values of the different community metrics for this case can be read from the corresponding non-evolving result indicated by the color map.
Along this curve, in contrast to the cases in which $R$ is independent of $D_v$, we can see, for instance, that the mean predation rate per prey is always close to its maximum value (Fig. S2B and Fig.~\ref{fig:catchingrate}) and, as a consequence, interspecific mixing is close to its minimum (see Figs. S2A and S4, which displays cuts of Fig. S2A at constant values of $D_v$ and includes an additional panel in which $D_v\neq D_p$). These constrains ultimately determine the behavior of the mean coexistence time and prey extinction probability as a function of $D_v$ (see Figs.~S2D and S2E), which correspond to the $\sigma_\mu = 0$ case in Fig.~\ref{fig:coexistence}.

%This correspondence between the non-evolutionary and the evolutionary model can be seen, for instance, in the behavior shown by both models for the interspecific mixing $\left<\mathcal{M}\right>$ (compare Fig. \ref{fig:mixing}A with Fig.~S2A), mean coexistence times, and prey extinction probability (compare Fig. \ref{fig:coexistence} with Figs. S2D and S2E) as a function of $D_v$ when mutation intensity is low. 
%\textbf{I THINK I CAN'T TOTALLY FOLLOW THE LAST PART OF THIS PARAGRAPH. I WOULD AVOID USING "READ OFF" AS TO ME IT MEANS "TO READ SOMETHING LOUDLY". IT PROBABLY HAS ANOTHER MEANING THAT FITS BETTER HERE BUT I DON'T KNOW IT AND ALSO COULDN'T FIND IT EASILY IN GOOGLE. ALSO, DO WE REALLY NEED ALL THE PANELS IN FIG. S2? IT SEEMS THAT WE ARE ONLY USING PANELS a, d and e. FINALLY, I WOULD AVOID THE USE OF THE EXPRESSION "PATH IN THE R-$D_v$ SPACE, BECAUSE TO ME THE TERM PATH HAS AN INTRINSIC MEANING OF "TRAJECTORY / TIME DEPENDENCY" WHEREAS HERE WE ARE JUST LOOKING AT FUNCTIONAL SHAPES (I AM MAKING THIS LAST SUGGESTION BECAUSE I HAD HARD TIME TO UNDERSTAND THIS MYSELF).}

Of course the approach above does not explain features occurring at high mutation intensity, such as the dependence of mixing on mutation (Fig. \ref{fig:deltamix}) or the maximum in $\mathcal{T}$ at intermediate mobility (Fig. \ref{fig:coexistence}A).  Nevertheless, additional insight for these results can still be extracted from the non-evolutionary simulations: a maximum in $\mathcal{T}$ similar to the one present in the evolutionary case for high mutations is obtained by varying $D_v$ at a sufficiently large, fixed value of $R$ (see supplementary Fig. S3). This suggests that the occurrence of this maximum is linked to the generation of large values of $R$ when mutation is large enough (see e.g. Fig. \ref{fig:HOMcase}B). To check this, we have performed simulations in which there is a fixed distribution of perceptual ranges $\rho(R)$ that is not evolving. This is achieved by assigning to each newborn predator a value of $R$ sampled from that distribution, instead of being inherited from the parent with some mutation. We use for the fixed distribution a Gaussian with given mean value and standard deviation. In situations in which the mean value or the standard deviation is sufficiently large, we recover the maximum in the coexistence time profiles for intermediate values of the mobility (supplementary Fig. S3). This confirms that large values of $R$, being they imposed, produced by mutation or by non-evolving variability, are responsible for the enhanced coexistence times. We have also checked (not shown) that the dependence of $\beta$ on mobility is also qualitatively reproduced under this non-evolving scenario if large values of $R$ are present.

The exploration of the $R-D_v$ plane in the non-evolutionary case helps to understand how the outcome of the evolutionary dynamics (the selected value $R^\star$ for each value of the diffusion coefficients) affects different population and community-level properties. A full understanding of the whole eco-evolutionary process (Fig.~\ref{fig:ecoevo}) requires also to address the other branch of the feedback loop, i.e. to understand how the ecological variables favor a particular perceptual range $R$. As we saw before (Fig.~\ref{fig:evo-comp}A), prey clumping and species segregation drive $R^\star$ to values higher than in the homogeneous case, since this increases the number of prey individuals potentially detected by predators. At the same time, however, attacking efficiency decreases for larger perceptual ranges, so that a finite optimal perceptual range, $R^\star$, results from this trade-off. The specific balancing value of $R$ is determined by the spatial distribution of the individuals in a way that we now try to elucidate. A first observation to be taken into account is that the selected $R^\star$ is always close to the one that minimizes interspecific mixing (Figs. S2A and S4). Thus, the selected perceptual range is the one producing the largest segregation between predator and prey. This increases $R^\star$ from its value at large mobility, $R_h^\star$, towards larger values as mobility is decreased (see also Fig. \ref{fig:evo-comp}).

A mechanism behind the shift of $R^\star$ with respect to $R_h^\star$ can be identified by realizing that the predation rate per prey, $\left<c(R)/N_v\right>_p$, closely follows the shape of $\left<\mathcal{M}\right>$ in the $R-D_v$ plane. Indeed, the selected $R^\star(D_v)$ in the eco-evolutionary model closely follows the maximum of $\left<c(R)/N_v\right>_p$ for each $D_v$ (Fig. S2B). The shift in the maximum of this function towards higher values of $R$ for decreasing $D_v$ was already seen in Fig. \ref{fig:catchingrate}. Note that this shift is not present in the quantity $\left<c(R)\right>_p$ (Fig. S2C). The ideas of adaptive dynamics \cite{Geritz1998} may be used to investigate the role of the predation rate per prey in selecting the optimal $R^\star$. $R^\star$ will be an evolutionary stable strategy, and thus selected under weak mutation, if the expected fitness of a mutant with perceptual range $R$ in a resident population with perceptual range $R^\star$ has a local maximum when $R=R^\star$. The fitness is defined here as the average net growth rate of the mutant predator in the fixed environment determined by $R^\star$. This is $\left<c(R)-d\right>_p$, where the predation rate is $c(R)=M_v(R)E(R)$ with an exponential attacking-efficiency function $E(R)$. The number of prey individuals available to the predator can be written as $M_v(R)=v\int_0^R g_{pv}(r)2\pi r dr$, where $g_{pv}$ is the radial pair correlation function\cite{LawDieckmann2000} $g_{pv}(r)$ that gives the expected number of prey individuals at distance a $r$ from a predator, relative to a random distribution with density $v=N_v/L^2$. $v$ and the functional form of $g_{pv}$ will depend, in general, on the value of the perceptual range. In determining the evolutionary stable value $R^\star$, these quantities must be kept constant while maximizing $\left<c(R)-d\right>_p$ with respect to the remaining explicit dependence on $R$ (which is a quantity exclusive to the mutant). One way to achieve that is to maximize the predation rate per prey, $\left<c(R)/N_v\right>_p=\left<E(R) L^{-2}\int_0^R g_{pv}(r)2\pi r dr\right>_p$, so that the explicit dependence of the fitness on $v$ or $N_v$ disappears. There will be still a residual dependence of $g_{pv}$ on $R$ but our numerical results indicate that it is weak and that maximization of the predation rate per prey (the maxima in the curves of Fig. \ref{fig:catchingrate} or of Fig. S2B as a function of $R$ for fixed $D_v$) gives a good approximation for the evolutionarily-stable selected perceptual range $R^\star$.

\section*{Summary and Discussion}
\label{sec:discussion}

Using an individual-based model, we have investigated whether the evolutionary dynamics of predator perceptual ranges influences the stability of spatially-structured prey-predator communities. First, we studied how different levels of interspecies mixing arise due to limited mobility and the variability in the perceptual range introduced by the intensity of mutations. Second, we evaluated the consequences of the interplay between species mixing and the predator perceptual range in other community-level properties. Our results reveal the existence of an eco-evolutionary feedback between interspecies mixing and predators perception: species mixing selects a certain distribution of perceptual ranges; in turn, the distribution of perceptual ranges reshapes species spatial distributions due to predation (predators eliminate prey from their surroundings). More specifically, when species mobility is low, prey and predators form monospecific clusters due to reproductive correlations\cite{Young2001}, but segregate from each other due to predation. Therefore, prey often inhabit regions of the environment that are not visited by predators, which drives the evolution of larger perceptual ranges. Conversely, as mobility becomes higher, species mixing increases and shorter-range predation is favored.
At the community level, we show that this eco-evolutionary feedback strongly controls both community stability and diversity, characterized by the mean coexistence time and prey extinction probability. The average coexistence time is maximum when the interaction between species mixing and predator perceptual ranges yields a predation rate that is large enough to sustain the population of predators but low enough to avoid fast extinctions of prey. We have also connected these results with those obtained for a purely ecological model, in which perceptual ranges can either vary across the population or be a deterministic trait.

The mean coexistence time and species extinction probabilities provide important information about the diversity of the community at different scales~\cite{rossine2018eco}. Each realization of our stochastic model can be seen as an independent dynamics taking place within a metapopulation or patch. In our setup, because model realizations are independent from each other, these patches are isolated (not coupled by dispersal events) and constitute a ``non-equlibrium metapopulation''~\cite{harrison}. In this context, the mean coexistence time is a proxy for alpha (intra-patch) diversity, i.e. how long species coexist in each patch, whereas species extinction probabilities inform about the beta (inter-patch) diversity, i.e., how many patches are expected to be occupied by prey and how many by predators once one of the species has been eliminated. From a mathematical point of view, the fraction of patches in which prey and predators coexist at any time $t$ is given by $P(t) = \int_t^{\infty} p (t') dt' = 1 - \int_0^{t} p (t')dt'$, where $p(t)$ is the distribution of coexistence times. Supported by our numerical simulations, we can approximate $p(t)\simeq \mathcal{T}^{-1}e^{-t/\mathcal{T}}$ (except for coexistence times that are much smaller than the mean, see supplementary Fig. S5). Therefore, $P(t) \simeq 1 - e^{-t/\mathcal{T}}$. The fraction of patches occupied only by prey is given by $(1-\beta)(1-P(t))$, and the fraction of patches in which overexploitation has caused prey extinction is given by $\beta(1-P(t))$. Hence, the mean coexistence time, $\mathcal{T}$, and the prey extinction probability, $\beta$, quantify the diversity of the community at different spatiotemporal scales~\cite{abgdiversity,diversityEntropy,rossine2018eco}, and might serve as important guides for the design of ecosystem management protocols~\cite{beta,abgdiversity}.

Since we were interested in studying whether the spatial coupling between movement and perception could lead to an eco-evolutionary feedback when both processes occur at comparable time scales \cite{rapid,rapid2}, we kept all the characteristic time scales of the system fixed except those related to individual movement and the evolutionary dynamics of perceptual ranges. We used parameter values such that evolution is fast enough so that both the spatial distribution of individuals and the distribution of perceptual ranges relax to their stationary values in timescales much shorter than the characteristic time scales at which community-level processes occur, defined by the mean coexistence time. Under this condition, the long-time regime is well-defined and can be characterized by constant quantities. A sensitivity analysis on reproduction probabilities (supplementary Fig. S6) reveals that, as far as this relationship between time scales is maintained, both the existence of the eco-evolutionary feedback and its impact on community stability and diversity remain unaffected. This condition can be broken, for instance, if prey and predator birth-death rates are small, or too unbalanced, producing very short coexistence times (Fig. S6). Our results for the mean coexistence time are robust against changes in the source of individual-level trait variability, as individual-level trait variability affects mixing in a similar way. This work was motivated by and focused on the case in which variability is produced by mutation in the transmission of the trait. However, our results could also be relevant to cases in which variability arises from non-inheritable properties, such as body size, or individual internal state (level of hunger, attention...) \cite{Fagan2017}, and therefore do not introduce spatial correlations between trait values after reproduction. When sampling the perceptual range of newborn predators from an appropriated fixed distribution (being independent on the parents' trait), the results in Fig.~\ref{fig:coexistence} are qualitatively reproduced (Fig. S3). This implies that different processes that promote trait variability can control community coexistence.

Finally, although we have concentrated our investigation on a prey-predator dynamics, our results will more generally illuminate whether, and to which extent, the interplay between species spatial distributions and the range of ecological interactions may determine community-level properties. Therefore, our study opens a broad range of questions and directions for future research. First, we have limited to the case in which only one predator trait can evolve, whereas evolution of prey traits has been also shown to impact profoundly the population dynamics of both species in well-mixed settings \cite{Cortez2010}. A natural extension of our study would be to explore such scenario in a spatially-explicit framework as the one introduced here. More complex possibilities, such as the co-evolution of traits in both species, including the possibility of evolving mobility rates, could also lead to new population dynamics \cite{Cortez2014,norrstrom2006,getz2016}. More general evolutionary processes, such as arms race in phenotype space (red queen-like dynamics) instead of trait distributions reaching a stationary configuration~\cite{armrace} also deserve further investigation. Our model considers that individuals undergo clonal reproduction. Extending it to the case of sexual reproduction can have important consequences for the evolutionary dynamics. On the one hand, sexual reproduction can speed-up evolution by increasing genetic diversity. On the other hand, sexual reproduction introduces a new range of additional processes, such as the cost of finding a mate and exposure to sexually transmitted diseases, that could change our results~\cite{getz2015,melian2012,getz2016,mamunoz}. Finally, different movement models, such as L\'evy flights instead of Brownian motion, can modify both the optimal range of the interactions \cite{Martinez-Garcia2014} and the emergence of clusters of interacting individuals \cite{Heinsalu2010}, possibly leading to new community-level results. The existence of environmental features that could also affect the degree of mixing and its coupling with the range of interactions, such as the presence of external flows, would extend our results to a wider range of ecosystems in which the importance of rapid evolutionary processes has been already reported \cite{pigolotti,turbulence}.

\section*{Methods}
\label{sec:methods}

\subsection*{Model details}
\label{sec:model}

We propose an individual-based model for a prey-predator
community in which each individual is represented by a point
particle within a square environment of lateral length $L$ and
periodic boundary conditions. In addition, each predator has an
individual-specific perceptual range $R$. In this context, we
propose an eco-evolutionary framework whose updating rules can
be grouped into three different types of dynamical processes:
individual movement (spatial dynamics), population dynamics and
evolution with mutation (dynamics in trait space).

\begin{enumerate}

\item \textit{Individual movement.} We assume that both
    prey and predators follow independent two-dimensional
    Brownian motions with diffusion constants $D_v$ and
    $D_p$ respectively. We sample a turning angle for
    individual $i$, $\theta_i$, from a uniform distribution
    between $[0,2\pi)$, and a displacement, $\ell_i$,
    defined as the absolute value of a normal random
    variable with zero mean and variance proportional to
    the individual diffusion coefficient. Mathematically,
    this updating in the position vector $\mathbf x_i(t)$
    of each individual can be written as
\begin{equation}\label{positionupdate}
\mathbf x_i(t+\Delta t) = \mathbf x_i(t) + \ell_i\boldsymbol{\hat{\theta}}_i \quad \forall\, i\,  \in\,  \{1,2,\ldots,N_p+N_v\}\, ,
\end{equation}
where $\boldsymbol{\hat{\theta}}_i = (\cos \theta_i, \sin
\theta_i)$ is the unitary random-direction vector and
$\ell_i$ is the length of the displacement, sampled from
the positive half of a normal distribution with second
moment $\bar{\ell_i}^2 = 2D_\alpha\Delta t$. The subscript
$\alpha =\lbrace v,p \rbrace$ refers either to prey or
predators, and $\Delta t$ is the simulation time step
defined below.

\item \textit{Population dynamics.} The number of prey
    individuals and predators, $N_v$ and $N_p$
    respectively, can change at every time step due to prey
    reproduction, predator death or predation. Predator
    death and prey reproduction occur with constant rates
    $d$ and $r$ respectively. Predation involves the
    encounter between one predator and one prey and
    therefore predation rates depend on predator-specific
    perceptual ranges, $R$, and the number of available
    prey individuals within it. A predator with perceptual
    range $R$ may thus successfully catch one of the
    accessible prey individuals and eliminate it (prey
    death) with rate $c(R)$. After prey elimination, the
    predator reproduces with probability $b$, which leads
    to a new individual at its position. The newborn
    individual inherits the parental perceptual range, $R$,
    plus a random contribution due to mutation. These three
    events can be written in the form of a set of
    biological reactions for prey, $V$, and predators, $P$,
\begin{align}
\label{reactions}
V &\xrightarrow{r} V + V \, , \notag\\
P &\xrightarrow{d} \emptyset \, , \\
P + V &\xrightarrow{c(R)}  \begin{cases}
      P + \tilde{P} & \text{with probability }b \\
      P & \text{with probability }1-b\, ,
   \end{cases} \notag
\end{align}
where we have added the notation $\tilde{P}$ to indicate
the variability in perceptual range inheritance due to
mutations (see Eq.~(\ref{mutation}) below).

A key step in our model is the definition of the predation rate, since it determines the interaction
between species and links predator perceptual ranges and
their reproductive success. The total predation rate of a
predator of perceptual range $R$, $c(R)$, is equal to the
number of prey individuals available within the predator
perceptual range, $M_v(R)$, multiplied by the attacking
efficiency, $E(R)$:
\begin{equation}
c(R) = E(R)M_v(R)\,.
\label{catchingrate}
\end{equation}
Since $M_v(R)$ is a monotonically increasing function of
$R$, $E$ must decrease sufficiently fast with $R$ in order
to bound the evolutionary dynamics in the perceptual range
and prevent the evolution of unrealistic infinite
perception. We write the attacking efficiency as
\begin{equation}
\label{eff}
E(R) = c_0 f(R)\, ,
\end{equation}
where $c_0$ is the maximum efficiency and $f(R)$ is the
dimensionless predator efficiency function. $f(R)$ is
considered to be a monotonically decreasing function of $R$
such that $f(0)=1$ and $f(R\rightarrow\infty) \rightarrow
0$. More specifically, since $M_v(R)$ grows as $R^2$ if
prey is homogeneously distributed, we use
$f(R)=\exp(-R/R_c)$ so its decay attenuates the growth of
$M_v(R)$ and $c(R)\rightarrow 0$ when $R\rightarrow\infty$.
$R_c$ fixes the characteristic spatial scale for the decay
of the predation efficiency with the perceptual range. This
simple form incorporates the essential ingredients to model
the trade-off, but in general, to account for effects that
might arise in the vanishing and large perceptual range
limits (for example a plateau with constant value of
efficiency for small $R$), forms beyond the exponential can
be implemented~\cite{getz1996}.

Given our particular choices, in the limit in which prey is
homogeneously distributed in space, the predation rate is
maximum at an intermediate perceptual range $R^\star_h
\equiv 2 R_c$. This optimal value $R^\star_h$ will be used
as a reference in our results. In addition, its existence
agrees with previous studies showing that, when information
can be gathered over long distance, foraging performance is
optimal for intermediate perceptual ranges
\cite{Martinez-Garcia2017b,Martinez-Garcia2013,
Martinez-Garcia2014, Fagan2017}. The trade-off between
perception and predation efficiency can also be motivated
by the case of flying predators. The area where prey can be
detected is generally increased by a higher flight
altitude. However, efficiency in the attack may
simultaneously be reduced if initiated from such a large
height. We remark that our predator mobility model only
describes the searching-for-prey phase. The attack process
is assumed to be instantaneous.
% described by the diffusion coefficient $D_p$ refers to the predator motion while searching. (I SUGGEST REMOVING THIS PART BECAUSE WE HAVE NOT INTRODUCED DIFFUSIONS YET).

\item \textit{Predator reproduction with mutation.} Each
    predation event is followed by the possible
    reproduction of the predator, occurring with
    probability $b$. Besides inheriting the parental
    position, the newborn individual also receives the
    parental perceptual range, $R$, but with an added
    random perturbation, $\xi_\mu$, that accounts for
    mutations. Therefore,
\begin{equation}
\label{mutation}
\tilde R = R + \xi_\mu\, .
\end{equation}
$\xi_\mu$ is a zero-mean Gaussian variable whose variance,
$\sigma_\mu^2$, regulates the intensity of the mutations.
In order to avoid perceptual ranges that exceed system size
or are negative, mutations leading to $R<0$ or $R>L/2$ are
rejected. %%
%We also do not consider any complexity in the genotype-phenotype map, so that we consider the phenotypic trait $R$ to be directly determined by the parental one and the mutations.\color{black}
\end{enumerate}

\subsection*{Model implementation: the Gillespie algorithm}

We implement the model stochastic birth-death dynamics
(processes 2 and 3 above) following the Gillespie algorithm
\cite{gillespie1977,gillespie1,gillespie2}. First, we compute
the total event rate $g = r N_v + \sum_{i=1}^{N_p} [c(R_i) +
d]$, summing each prey reproduction rate and each predator
catching and death rates. Then, the simulation time-step is set
to $\Delta t = \zeta \tau$, where $\zeta$ is an exponentially
distributed random variable with unit mean and $\tau \equiv
1/g$ is the average time to the next demographic event. At each
iteration, a population dynamics event will occur chosen from
all ($N_v + 2 N_p$) possible events (i.e. choosing one of the
$N_v$ prey individuals for reproduction, or one of the $N_p$
predators for death, or one of them for a catching event). The
probability of choosing a particular event is proportional to
its relative contribution to the total rate $g$, i.e. each prey
has probability $r/g$ of generating a new prey, and each
predator has probability $d/g$ of dying and a probability
$c(R_i)/g$ to catch a prey. If predator death or prey
reproduction occur, we simply remove or generate a new
individual at parents' position, respectively. And finally, if
predation occurs, a prey randomly chosen within the perceptual
range of the selected predator $i$ (which was chosen
proportionally to its rate $c(R_i)$) dies and, with probability
$b$, a new predator is generated at the same location as the
initial predator, with value of the perceptual range obtained
from Eq.~(\ref{mutation}).

For simplicity, we fix across this paper the parameter values
$r=d=c_0=b=R_c=1$ and focus our study on the role of the
diffusion coefficients $D_p$ and $D_p$, and the mutation
intensity, $\sigma^2_{\mu}$, on determining the dynamics of the
community (for computational convenience we use also different
values of system size $L$). The impact of changing some
parameter to other values is briefly addressed in the
\textit{Summary and Discussion} section.

\subsection*{Mixing measures}
\label{sec:mixing}

In order to quantify the spatial arrangement of the species, we
define measures of mixing. A possible way to proceed is to use
the Shannon index or entropy, which has been applied to measure
species diversity, racial, social or economic segregation on
human population and as a clustering
measure~\cite{diversityEntropy,entropyindex,cristobalentropy}.
Based on these previous approaches, we propose a modification
described below.

As usual, we start regularly partitioning the system in $m$ square boxes with size $\delta x = L/\sqrt{m}$ and
obtaining for each box $i$ the \emph{entropy index} $s_i$~\cite{entropyindex}, given by
\begin{equation}
\label{mixbox}
s_i = - f_p^{(i)} \ln f_p^{(i)} - f_v^{(i)} \ln f_v^{(i)}\, ,
\end{equation}
where $f_p^{(i)}$ ($f_v^{(i)}$) is the fraction of predators
(prey) inside box $i$, i.e. $f_q^{(i)} =
N_q^{(i)}/[N_p^{(i)}+N_v^{(i)}]$ with $q=p,n$ and $N_p^{(i)}$,
$N_v^{(i)}$ the numbers of predators and prey individuals in
that box, respectively. In terms of Eq.~(\ref{mixbox}),
prey-predator mixing is maximum when there is half of each type
in the box, yielding $s_i = -\ln 1/2 = \ln 2$. Unbalancing the
proportions of the two types in the box reduces $s_i$. If a box
contains only predators or prey, $s_i=0$, indicating perfect
segregation. Finally, we define a whole-system prey-predator
mixing measure by averaging the values $s_i$ in the different
boxes, each one weighted by its local
population~\cite{entropyindex},
\begin{equation}
\langle \mathcal{M} \rangle_m \equiv  \sum_{i=1}^m \frac{N^{(i)}}{N}s_i\, ,
\label{avgmbox}
\end{equation}
being $N^{(i)} = N_p^{(i)}+N_v^{(i)}$ the total box population
and $N=N_v+N_p$ the total population. To really characterize
the lack of inhomogeneity arising from interactions and
mobility, one should compare the value of $\langle \mathcal{M}
\rangle_m$ with the value $\overline{\mathcal{M}}$ that would
be obtained by randomly locating the same numbers of predators
and prey individuals, $N_p$ and $N_v$ among the different
boxes. At this point, approximations for
$\overline{\mathcal{M}}$ which are only appropriate if the
number of individuals is large have been typically used. In our
case, since predator and prey populations have large
fluctuations, it is necessary to give a more precise
estimation. In a brute force manner, one can obtain
computationally the mixing measure for the random distribution
simply by distributing randomly in the $m$ spatial boxes the
$N_v$ prey individuals and $N_p$ predators and averaging the
corresponding results of Eq.~(\ref{avgmbox}) over many runs. On
the other hand, this can be done analytically since we known
that, for random spatial distribution, the number of
individuals $n_q$ of type $q$ ($=p, v$) in each box would obey
a binomial distribution $B(n_q,N_q)$, where $N_q$ is the total
particle number in the system. We have that $B(n_q,N_q) =
\binom{N_q}{n_q}
(\frac{1}{m})^{n_q}(1-\frac{1}{m})^{N_q-n_q}\,$. Then,
Eq.~(\ref{avgmbox}) for randomly mixed individuals becomes
\begin{equation}
\overline{\mathcal{M}}  \equiv  \sum^{N_v}_{n_v=0} \sum^{N_p}_{n_p=0} B(n_v,N_v)B(n_p,N_p) \frac{n_v+n_p}{N_v+N_p} s(n_v,n_p)\, ,
\end{equation}
with $s$ the entropy index in a box containing $n_v$ prey
individuals and $n_p$ predators, as defined in
Eq.~(\ref{mixbox}).

Finally, a suitable measure of prey-predator mixing that
characterizes spatial structure from the well-mixed case
($\mathcal{M} = 1$) to full segregation ($\mathcal{M} = 0$) is
given by
\begin{equation}
\mathcal{M} \equiv \frac{\langle \mathcal{M} \rangle_m}{\overline{\mathcal{M}}}\, .
\end{equation}

Also, we can define an analogous measure for each species'
spatial distribution separately, which can be interpreted as a
degree of clustering\cite{cristobalentropy},
\begin{equation}
\mathcal{M}_v = -\frac{1}{\overline{\mathcal{M}}_v}\sum_i^m \frac{(N_v^{(i)}/N_v)\ln (N_v^{(i)}/N_v)}{m}\, ,
\end{equation}
and
\begin{equation}
\mathcal{M}_p = -\frac{1}{\overline{\mathcal{M}}_p}\sum_i^m \frac{(N_p^{(i)}/N_p)\ln (N_p^{(i)}/N_p)}{m}\, ,
\end{equation}
for prey and predators respectively, where
\begin{equation}
\overline{\mathcal{M}}_v = \sum^{N_v}_{n_v=0} B(n_v,N_v)[ (n_v/N_v)\ln (n_v/N_v)]\, , \quad  \overline{\mathcal{M}}_p = \sum^{N_p}_{n_p=0} B(n_p,N_p)[ (n_p/N_p)\ln (n_p/N_p)]\, .
\end{equation}
For $\mathcal{M}_v$ or $\mathcal{M}_p=1$, the corresponding
species is well spread around the domain. Smaller values
indicate clustering of the individuals.

The mixing measures are certainly affected by the size of the
box $\delta x$ used, which should be tuned to obtain maximum
sensibility to the spatial distribution. For very large as well
as for very small box size, we see that different spatial
distributions become indistinguishable. For instance, for the
prey-predator mixing, if the box size is very large (of the
order of system size), we will find that the predators and prey
are well-mixed independently of the values of the diffusion
coefficients. On the other hand, if box size is very small it
will be either occupied by a single predator or prey, if not
empty, indicating segregation independently on the individual
mobility. In supplementary Fig. S7, we show how the mixing
measure changes with box size $\delta x$ and system size $L$
for low mobility ($D_v=D_p=0.1$), which produces a highly
heterogeneous spatial distribution. We identify that for
$\delta x \simeq 2$ the maximum sensitivity with respect to the
diffusion coefficients is attained (for $L=10$). We used
$\delta x = 2$ in our results, being a suitable scale since it
is also of the order of the typical values of the perceptual
range attained under evolution. Regarding the system size, we
found only weak variations in the mixing measures, which are
shown in the inset of Fig. S7.

\section*{Data availability}
No datasets were generated or analysed during the current study.

\section*{Acknowledgments}
This work is partially funded by the Gordon and Betty Moore
Foundation through Grant GBMF2550.06 to RMG, and by the Spanish
Research Agency, through grant ESOTECOS FIS2015-63628-C2-1-R
(AEI/FEDER, EU) and grant MDM-2017-0711 from the Maria de
Maeztu Program for units of Excellence in R\&D (EHC, CL and
EH-G).

\section*{Author contributions}

E.H.C. performed the research and analyzed data; E.H.C and
R.M-G. drafted the manuscript; E.H.C., R.M-G., C.L., E.H-G.
designed and planned the study; E.H.C., R.M-G., C.L., E.H-G.
reviewed and edited the final version of the manuscript.

\section*{Additional information}

\textbf{Competing interests}. The authors declare no competing interests.

\end{document}